\documentclass[useAMS,usenatbib]{mn2e}
\usepackage{amssymb}

\usepackage{graphicx}
\usepackage{float}

\title[Exploring masses \& abundances of red giants]{Exploring masses and CNO surface abundances of red giant stars}

\author[G. M. Halabi \& M. El Eid]{Ghina M. Halabi$^{1}$\thanks{E-mail: gm29@aub.edu.lb}, Mounib El Eid$^{1}$\\
$^{1}$Department of Physics, American University of Beirut, Bliss Street 11-0236, Beirut 1107 2020, Lebanon\\
}
\begin{document}
\bibliographystyle{plainnat}

\date{Accepted: May 2015. Received ; in original form }

\pagerange{\pageref{firstpage}--\pageref{lastpage}} \pubyear{2015}

\maketitle

\label{firstpage}

\begin{abstract}
A grid of evolutionary sequences of stars in the mass range $1.2$-$7$M$_{\odot}$, with solar-like initial composition is presented. We focus on
this mass range in order to estimate the masses and calculate the CNO surface
abundances of a sample of observed red giants. The stellar models are
calculated from the zero-age main sequence till the early asymptotic giant
branch (AGB) phase. Stars of M $\leqslant$ $2.2$M$_{\odot}$ are evolved
through the core helium flash. In this work, an approach is adopted that
improves the mass determination of an observed sample of 21 RGB and early
AGB stars. This approach is based on comparing the observationally derived
effective temperatures and absolute magnitudes with the calculated values
based on our evolutionary tracks in the Hertzsprung-Russell diagram. A more reliable
determination of the stellar masses is achieved by using evolutionary tracks
extended to the range of observation. In addition, the predicted CNO surface
abundances are compared to the observationally inferred values in order to
show how far standard evolutionary calculation can be used to interpret
available observations and to illustrate the role of convective mixing. We
find that extra mixing beyond the convective boundary determined by the
Schwarzschild criterion is needed to explain the observational oxygen isotopic
ratios in low mass stars. The effect of recent determinations of proton
capture reactions and their uncertainties on the $^{16}$O$/^{17}$O and $^{14}%
$N$/^{15}$N ratios is also shown$.$ It is found that the $^{14}$N$($%
p$,\gamma)^{15}$O reaction is important for predicting the $^{14}$N$/^{15}$N
ratio in red giants.
\end{abstract}


\section{Introduction}
After the main sequence evolutionary phase, stars evolve to the red giant
branch (RGB). This evolution is initiated by the ignition of shell H-burning
surrounding the He core, whose energy flux causes the envelope to expand and
the star evolves to the RGB. The expansion increases the opacity and leads to
the development of a deep convective envelope. This is the first dredge up
event (FDUP), as convection mixes up the products of H-burning to the surface
altering the surface composition of the star.

In the mass range ($4$-$7)$M$_{\odot}$, stars exhibit blue loops at the
beginning of core He burning (see \citet{hal12}, and references therein). The
main phase of core He-burning is completed before the track evolves back to
the RGB. This leads again to the deepening of envelope convection. For solar
metallicity stars of M $\gtrsim$4M$_{\odot}$, a second dredge up (SDUP) can
reach deeper regions, which leads to further changes in the surface abundances.

This work uses observations obtained for a sample of red giants by
\citet{tsu08}, hereafter Tsuji08, in order to achieve two goals: (a) to
estimate the masses of these giants by matching their observationally derived
effective temperatures and bolometric magnitudes to the values obtained from
the evolutionary tracks in the HR diagram. This is possible since the stars
are not pulsating Mira variables (see Section 3.1 for details). We are able to
improve the mass determination done by Tsuji08 by using more extended
evolutionary tracks to avoid the extrapolation that he partially relied on to
determine the mass of some giants, (b) to compare the calculated CNO
abundances of these models to those inferred from observations in the light of
recent determinations of key reaction rates.

A large body of observational data is available for the surface CNO abundances
in RGB stars
\citep{lam81,har84a,har84b,har88,lam86,gil91,tsu91,cha94,tsu08,tau10,pia11}.
These data provide a powerful tool to get insight into the internal structure
of evolved stars. In particular, comparing the predicted oxygen surface
abundances to the observed data is useful to improve the treatment of
convective mixing in the stellar interiors (see Section 3.2.2). Our results
suggest the need for extra mixing below the edge of the convective envelope as
determined by the Schwarzschild criterion in order to achieve a better
agreement with observations. This extra mixing has been suggested in several
investigations in connection with the evolution of field giants
\citep{cha98,gra00}, open clusters \citep{luc94,tau00,tau05}, globular
clusters \citep{she03,pil03,rec07,den14} as well as to explain isotopic ratios
in pre-solar grains \citep{pal11,pal13,bus14}. The abundance profile of
$^{17}$O of particular interest. This is because this isotope is produced by
the ON-cycle which requires higher temperatures than the CN-cycle. Therefore,
the $^{17}$O profile exhibits a steep gradient within the central region of
the star, at the position of maximum convective penetration (shown later in
Fig. \ref{fig9}). This renders the $^{17}$O surface abundance sensitive to the
depth of convective mixing, stellar mass and to the nuclear reaction rates
involved in the CNO\ cycle \citep{eid94}. We show in Section 3.2.2 how the
$^{16}$O$/^{17}$O ratio is useful to constrain this extra mixing. The low
surface carbon isotopic ratios in low mass stars however, can not be explained
by our extra-mixing treatment. Other non-standard mixing mechanisms may need
to be invoked as shown in Section 3.2. \citet{ram14} also discuss carbon
isotopic ratios in AGB\ stars in connection with their recent estimations of
circumstellar $^{12}$CO$/^{13}$CO abundance ratios based on radiative transfer
analysis of radio line emission observations.

Concerning the nuclear reaction network, there has been extensive experimental
work recently on updated determinations of major reaction rates, including
those of the CNO cycle. One set of rates that we use \citep{ser14} was
obtained with the Trojan Horse method \citep{lac10}, which is an indirect
technique that is able to provide more reliable reaction rate cross-sections
at low temperatures where measurements in the astrophysical energy range are
available. Other sets, obtained by \citet{sa13} and \citet{il10} are evaluated
based on Monte Carlo techniques \citep{lon10}. This method provides a median
rate which -under certain conditions- resembles the commonly referred to
\textquotedblleft recommended\textquotedblright\ rate, as well as a low rate
and a high rate which, unlike the \textquotedblleft upper\textquotedblright%
\ and \textquotedblleft lower\textquotedblright\ limits of classical rates,
have a well-defined statistical meaning. The effect of these recent
determinations of the proton capture reactions including the $^{14}$%
N$($p$,\gamma)^{15}$O rate on the isotopic ratios $^{16}$O$/^{17}$O and
$^{14}$N$/^{15}$N is investigated$.$

This paper is organized as follows. Section 2 describes the calculations and
provides a summary of the observational data used in this work. The results
are presented in Section 3. Mass determinations and comparison with previous
works are provided in Section 3.1. In Section 3.2, we describe the surface
abundance profiles and the effects of extra mixing. The effect of nuclear
reaction rates is discussed in Section 3.3. Concluding remarks are given in
Section 4.
\section{Model Calculations}

The evolutionary sequences presented in this work are obtained using the
stellar evolution code HYADES as described in \citet{eid09}, with the recent
modifications outlined in \citet{hal12}. This code is a one-dimensional
implicit Lagrangian code based on a hydrodynamical method which solves the
stellar structure equations on an adaptive grid. Mass-loss is included using
semi-empirical rates adjusted to the global parameters of the star. The rates
are used according to a Mira pulsation period ($P$) \citep{vas92}. For $P$$<$100 d, \citet{rei75} mass-loss rate is used, with\ $\eta$ =$1$. For
$100\leqslant$P$<$500 d we use a more effective rate according to \citet{bow88}. For
$P\geqslant500$ d, the superwind mass-loss rate during the AGB is used as
suggested by \citet{vas92}.

\subsection{Analyzing the overshooting region}

In the context of the Mixing Length Theory (MLT), the extension of a
convective zone is determined by the Schwarzschild criterion, that is when
$\nabla_{\textnormal{rad }}>\nabla_{\textnormal{ad}},$ where $\nabla_{\textnormal{rad }}$and
$\nabla_{\textnormal{ad}}$ are the radiative and adiabatic temperature gradients, respectively.

Within this framework, a long-standing issue is to fix the boundary of the
convective zone. In the local description of the MLT, mixing beyond the
Schwarzschild boundary is introduced in a parameterized way. According the
multi-dimensional hydrodynamic simulations by \citet{fre96}, a local
description of an extra mixing (or overshooting) may be introduced in terms of
an exponentially decaying diffusion coefficient:
\begin{equation}
\textnormal{D}(\textnormal{z})=\textnormal{D}_{o}e^{\frac{-2\textnormal{z}}{fH_{p}}}\textnormal{ \ \ }
\label{diff}%
\end{equation}
where z$=\left\vert r_{\textnormal{boundary}}-r\right\vert $ is the overshoot
distance, D$_{o}$ is the diffusion coefficient at the boundary of the
convective envelope obtained from the Mixing Length Theory (see
\citet{lan85}), H$_{\textnormal{p}}$ is the pressure scale height and $f$ is a free
parameter which is a measure of the efficiency of this extra partial mixing.
It is clear from Eq. \ref{diff} that for smaller values of $f,$ D has a
steeper profile, or equivalently less extra mixing. As $f$ increases, this
extra mixing extends further beyond the formal convective boundary. The
numerical simulations by \citet{fre96} find $f$ =0.25 $\pm$\ 0.05 and 1.0
$\pm$\ 0.1 for A-stars and DA white dwarfs, respectively. We will use the
observationally inferred oxygen isotopic ratio in red giants to constrain the
value of $f$ (see Section 3.2).

The mixing of chemical elements is achieved by solving the diffusion equation
given by:%

\begin{equation}
\frac{\textnormal{dX}_{i}}{\textnormal{dt}}=\frac{\partial}{\partial M_{r}}[(4\pi
r^{2}\rho)^{2}\textnormal{D}\frac{\partial X_{i}}{\partial M_{r}}] \label{dif}%
\end{equation}
where $r$ is the radius, $\rho$ is the mass density and D is\ the diffusion
coefficient given by Eq. \ref{diff} when used in the overshoot region,
otherwise it is equal to D$_{o}$ in a convective zone according to the
Schwarzschild criterion.%

\begin{figure*}
\centering
 \includegraphics[scale=0.416]{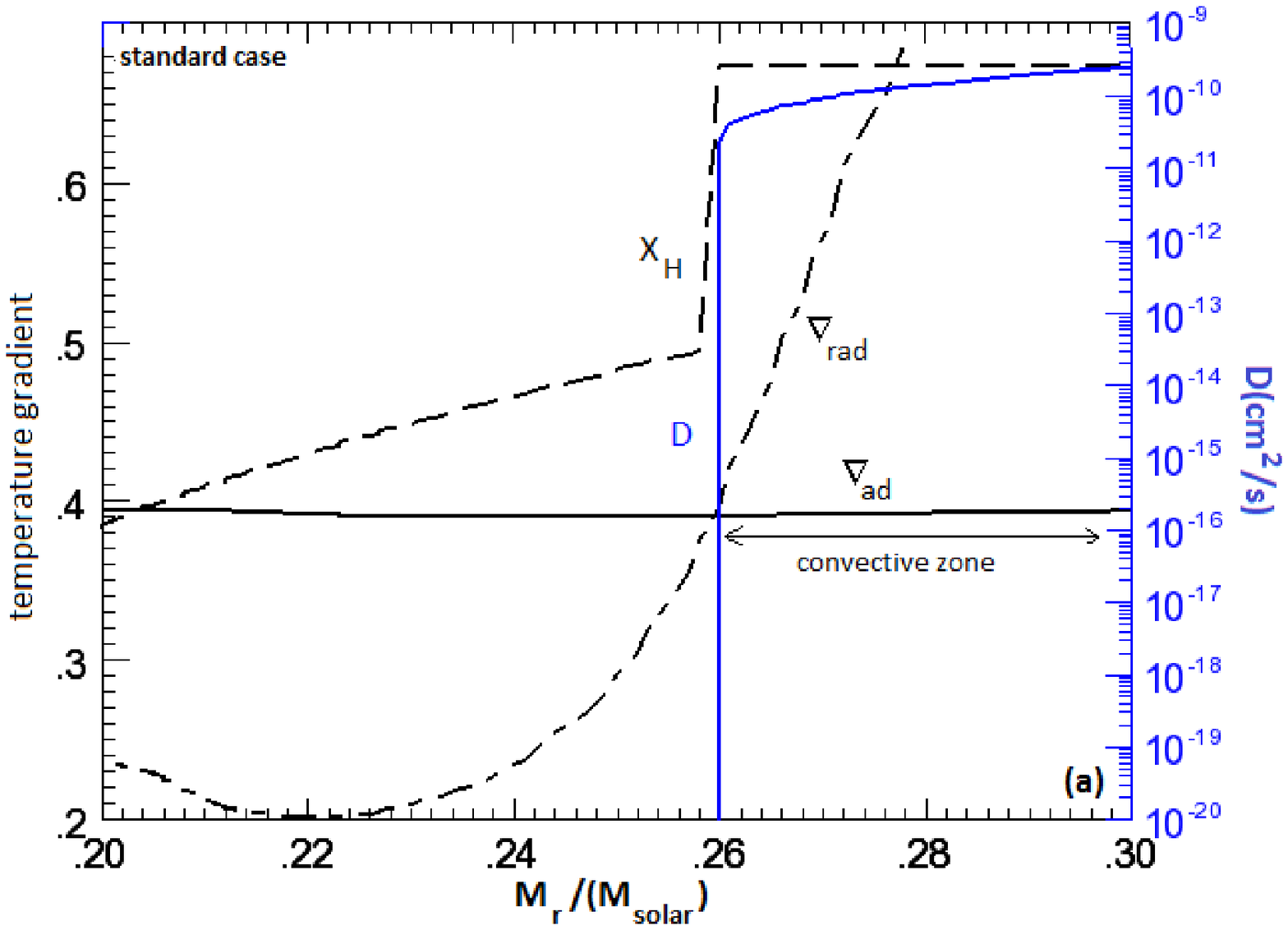} 
 \includegraphics[scale=0.415]{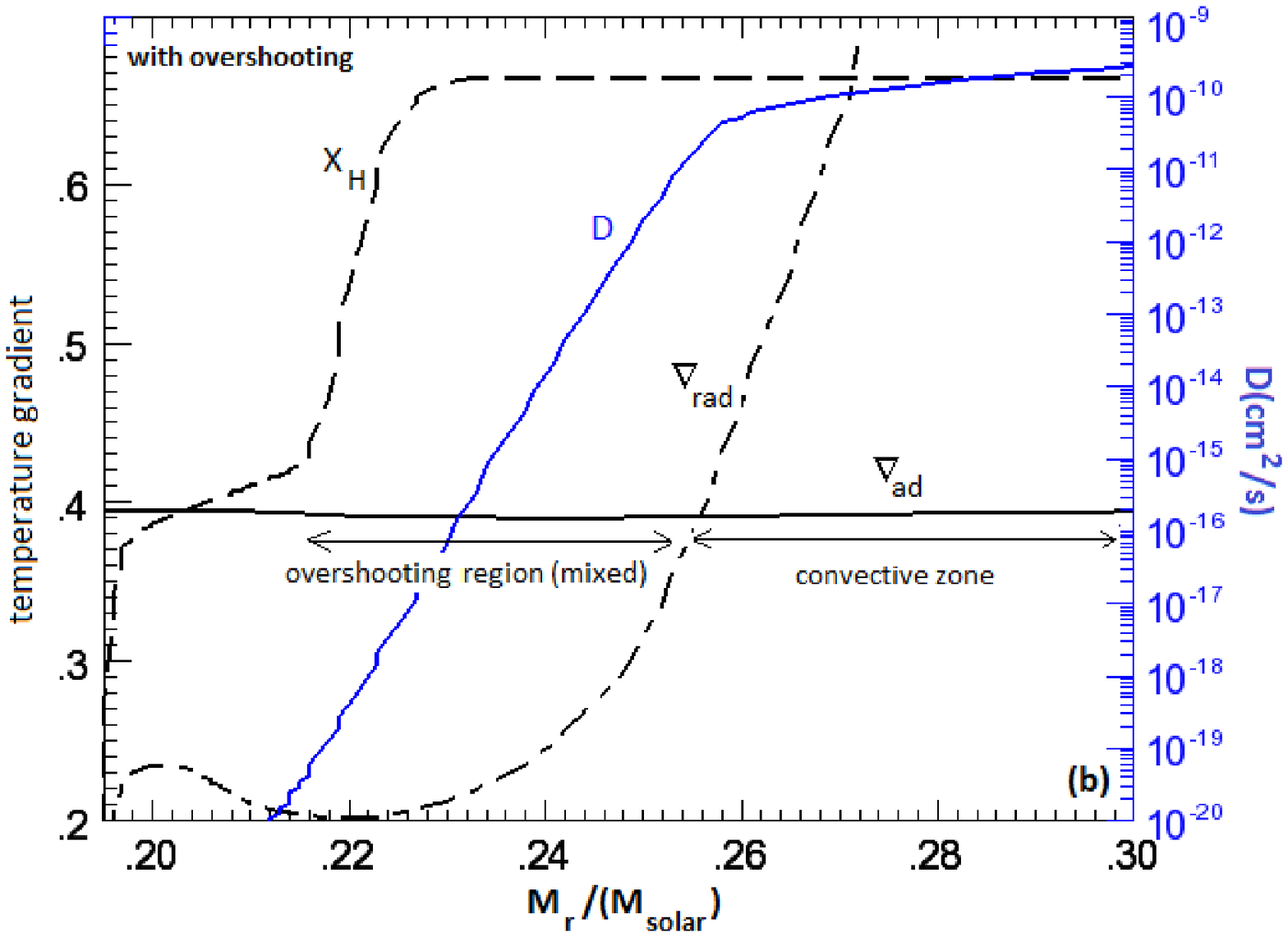}
 \includegraphics[scale=0.415]{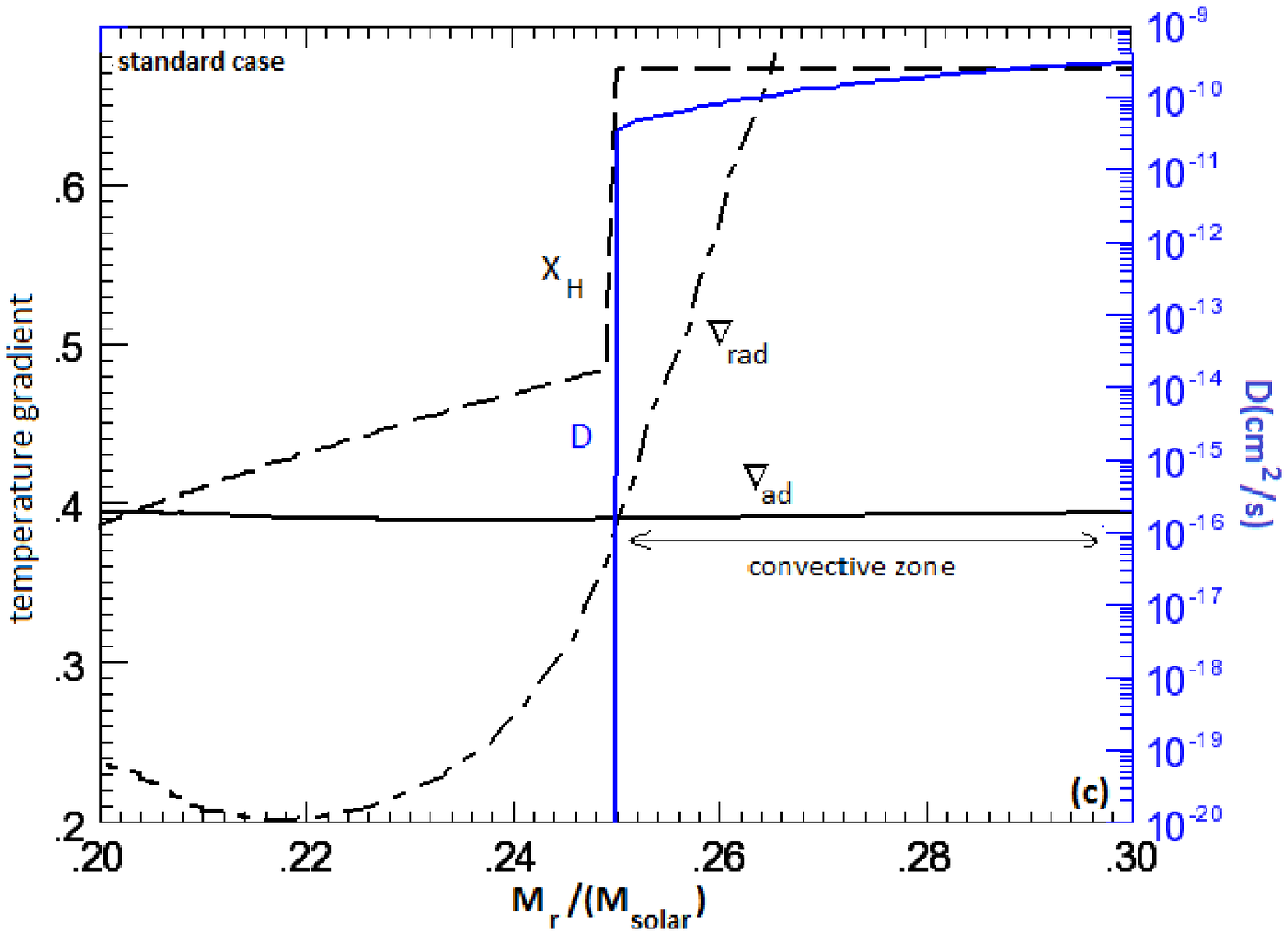}
 \includegraphics[scale=0.415]{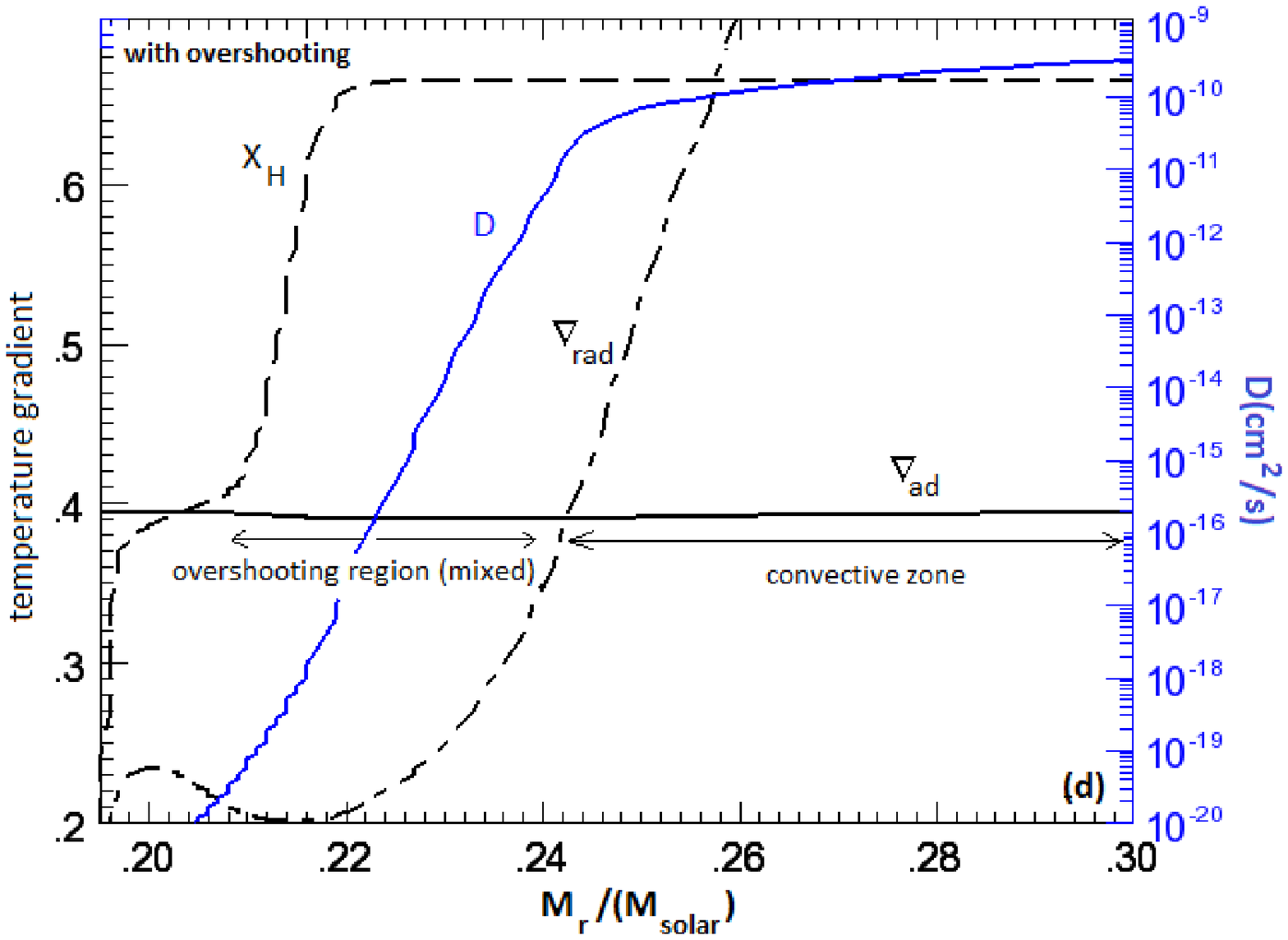}
\caption{Physical quantities related to the condition of convection at the bottom of the convective envelope in a 1.2M$_{\odot}$ star during FDUP, for (a) the standard stellar model and (b) the model with overshooting using f=0.125. (c) and (d) are the same as (a) and (b), respectively, but 3$\times10^{5}$ years later.  $\nabla_{\textnormal{rad }}$ and $\nabla_{\textnormal{ad }}$
are the radiative and adiabatic temperature gradients, respectively. Also shown are the hydrogen profile (left scale) and diffusion coefficient profile (right scale).}\label{grad}%
\end{figure*}%

To illustrate the effect of the treatment of mixing described above, Fig.
\ref{grad} shows the behavior of $\nabla_{\textnormal{rad }},$ $\nabla_{\textnormal{ad }}$
as well as the profiles of the diffusion coefficient and hydrogen
(X$_{\textnormal{H}}$) as a function of interior mass. This is done for a
1.2M$_{\odot}$ model during the first dredge up phase (FDUP) after the star
has evolved to the red giant branch.

The left-hand panel of Fig. \ref{grad} shows the extension of the convective
envelope in the standard models, i.e. without overshooting where the abrupt
drop of the diffusion coefficient D is visible along with the vertical drop of
X$_{\textnormal{H}}.$ The right panel shows the behavior with overshooting. It is
seen that in the latter case, mixing is extended into the radiative region
where $\nabla_{\textnormal{rad }}<\nabla_{\textnormal{ad}}$. The lower panels show the
profiles at the next time-step, 3$\times10^{5}$ years later. In both cases,
the convective envelope deepens in mass as FDUP proceeds, but it is deeper in
the case with overshooting, where the change in composition alters the opacity
so that $\nabla_{\textnormal{rad }}>\nabla_{\textnormal{ad}}$ becomes satisfied deeper in
mass. Therefore, overshooting does not only induce extra mixing but also
drives convective instability. Details of the present calculations with this
overshooting or extra diffusive mixing are provided in Section 3.2.2.

\subsection{Observational data}

A sample of red giant stars has been observed by Tsuji08 as given in Table
\ref{tab1}. The effective temperatures were obtained using the infrared flux
method \citep{bla80}, while the absolute bolometric magnitudes were determined
from the bolometric luminosities obtained by integrating the spectral energy
distributions and the Hipparcos parallaxes. The uncertainty on the effective
temperature is estimated to be $100$K, and the error on the bolometric
magnitude is mainly due to the error on the parallaxes. In the next section,
the effective temperatures and bolometric magnitudes are used to determine the
stellar masses of the observed giants using our evolutionary tracks, which
cover the whole range of observations.%

\begin{table}%
\caption{Spectral type, effective temperatures and bolometric magnitudes (Tsuji08).}%
\begin{tabular}
[c]{|l|l|l|l|}\hline
Object(BS/HD) & Spectral type & T$_{eff}$ (K) & M$_{\textnormal{bol}}$
(mag)\\\hline\hline
$\delta$ Vir $(4910)$ & M3III & $3643$ & $-2.4\pm\ 0.3$\\\hline
$\alpha$ Tau $(1457)$ & K5+III & $3874$ & $-1.7\pm\ 0.2$\\\hline
RRUMi $(5589)$ & M4.5III & $3397$ & $-3.4\pm\ 0.3$\\\hline
RZ Ari $(687)$ & M6III & $3341$ & $-3.5\pm\ 0.6$\\\hline
$\delta$ Oph $(6056)$ & M0.5III & $3790$ & $-2.2\pm\ 0.3$\\\hline
$\nu$ Vir $(4517)$ & M1III & $3812$ & $-2.2\pm\ 0.4$\\\hline
$\tau^{4}$ Eri $(1003)$ & M3+IIIa & $3712$ & $-2.9\pm\ 0.4$\\\hline
$10$ Dra $(5226)$ & M3.5III & $3730$ & $-2.9\pm\ 0.3$\\\hline
$\beta$ Peg $(8775)$ & M2.5II-III & $3606$ & $-3.3\pm\ 0.2$\\\hline
$30g$ Her $(6146)$ & M6-III & $3298$ & $-4.2\pm\ 0.4$\\\hline
$\sigma$ Lib $(5603)$ & M2.5III & $3596$ & $-3.4\pm\ 0.5$\\\hline
R Lyr $(7157)$ & M5III & $3313$ & $-4.3\pm\ 0.3$\\\hline
$\mu$ Gem $(2286)$ & M3III & $3643$ & $-3.3\pm\ 0.3$\\\hline
OP Her $(6702)$ & M5II & $3325$ & $-4.4\pm\ 0.8$\\\hline
$\rho$ Per $(921)$ & M4II & $3523$ & $-4.1\pm\ 0.4$\\\hline
$\alpha$ Cet $(911)$ & M1.5IIIa & $3909$ & $-3.2\pm\ 0.3$\\\hline
$\lambda$ Aqr $(8698)$ & M2.5III & $3852$ & $-3.4\pm\ 0.7$\\\hline
XY Lyr $(7009)$ & M5II & $3300$ & $-5.1\pm\ 1.1$\\\hline
$\delta^{2}$ Lyr $(7139)$ & M4II & $3420$ & $-5.5\pm\ 0.8$\\\hline
$\alpha$ Her $(6406)$ & M5Ib-II & $3293$ & $-5.8\pm\ 1.6$\\\hline
BS6861$(6861)$ & M4 & $3600$ & $-5.2\pm\ 2.0$\\\hline
\end{tabular}
\label{tab1}%
\end{table}%

\section{Evolutionary Results}

Evolutionary sequences of stars in the mass range (1.2-7.3)M$_{\odot}$ are
calculated from the zero-age main sequence till the early AGB phase. Stars of
M $\geqslant$ $3$M$_{\odot}$ exhibit blue loops starting at the onset of
central helium burning, which become more extended for stars of mass $\gtrsim$
$4$M$_{\odot}.$ A detailed discussion on this evolutionary phase has been
presented in \citet{hal12}. Stars of masses $\leqslant2.2$M$_{\odot}$ and
solar-like initial composition evolve through the core He-flash
\citep{kip90,moc10,bil12}. We find that for M $<$ $2$M$_{\odot}$, the
helium flash starts off-center owing to the cooling via the plasma and photo
neutrino energy losses. In the mass range $2\leqslant$M/M$_{\odot}%
\leqslant2.2$ the helium flash starts at the center since these stars evolve
at relatively lower central densities so the cooling by neutrino energy losses
is less efficient.

The core helium flash requires very short time steps to accommodate the
rapidly changing variables. In our calculation, the time step is of the order
of less than a year during the core helium flash. For stars of masses in
excess of $2.2$M$_{\odot}$, no significant degeneracy effects occur and core
He-burning proceeds under hydrostatic conditions.

\subsection{Mass evaluation}

In the work by Tsuji08, the masses of the red giant stars listed in Table
\ref{tab1} are derived using the evolutionary tracks by \citet{cla04}. A main
difference between our code and that of \citet{cla04} is that the latter used
\citet{cau88} rates for the basic nuclear reactions in the network, while our
used reaction rates are updated according to the JINA\ REACLIB\ database
\citep{cyb10}. This is expected to introduce modifications on the evolutionary
tracks. Another difference is that Claret assumes core overshooting, with an
overshooting distance of 0.2H$_{\textnormal{p }}$ for the whole mass range and
ignores envelope overshooting. Core overshooting results in a bigger core mass
and affects the evolutionary tracks and the stellar lifetimes. In the sample
considered here, most of the stars are low-mass stars which have very small or
no convective core at all, thus applying the same amount of overshooting at
the convective core boundary would result in a large amount of mixing that
yields results which are inconsistent with observations \citep{woo01}. In
order to avoid this artifact, core overshooting is not included in our calculation.

Moreover, the tracks by \citet{cla04} are not extended enough, so that
extrapolations were needed at temperatures below $3200$K during the RGB phase
for about half of the stars in the sample studied by Tsuji08. In this
calculation, the stars are evolved until the early AGB phase without relying
on extrapolation. Moreover, mass-loss is included, and we report the masses of
the evolved stars rather than their initial masses, which improves the mass
determination, particularly for the higher masses where mass-loss becomes more effective.

\begin{figure*}
\begin{center}
\includegraphics[scale=0.9]{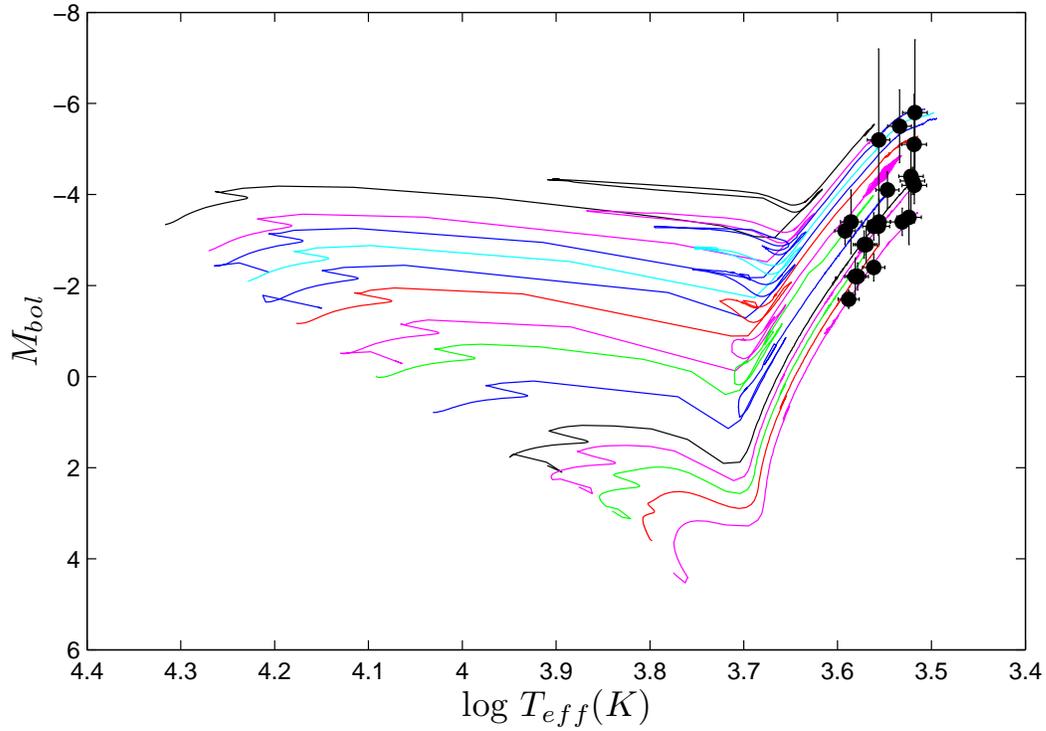}%
\caption{Bolometric magnitude (M$_{\textnormal{bol}})$ versus effective temperature
(T$_{eff})$ showing the present evolutionary tracks of stars of masses
(1.2-7)M$_{\odot}$. Also shown is the observed sample listed in Table
\ref{tab1}.}%
\label{fig5}%
\end{center}
\end{figure*}

Adopting the values of M$_{\textnormal{bol}}$ and T$_{eff}$ given in Table
\ref{tab1}, the evolutionary tracks shown in Fig. \ref{fig5} are used to
evaluate the stellar masses. It is important to note here that a direct
comparison of the theoretical temperature to that inferred observationally
wouldn't have been possible if the stars are pulsating Mira variables, in
which case a radius cannot be strictly defined and any comparison wouldn't
hold. Even the term effective temperature may become questionable for the very
evolved AGB stars featuring strong pulsations and mass loss rates
\citep{bas91,leb10}. However, in this sample, the stars are on the RGB or
early AGB phase, and thus, haven't yet experienced any thermal pulsations.
This allows a reliable comparison between the predicted effective temperatures
and the observationally inferred ones. The theoretical bolometric magnitude is
obtained using the well known relation: M$_{\textnormal{bol}}=4.75-2.5\log($%
L$/$L$_{\odot}).$

\begin{figure*}
\begin{center}
\includegraphics[scale=0.7]{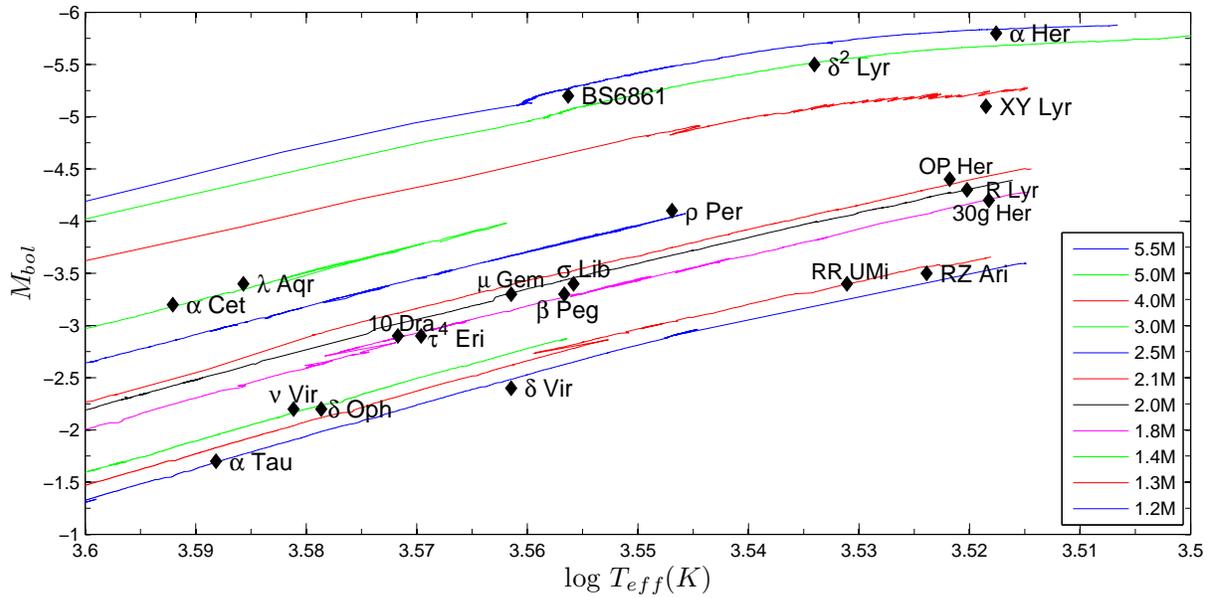}%
\caption{The evolutionary tracks together with the observed red giants listed
in Table \ref{tab1}. Note that the observed data points have errors as
indicated in Table \ref{tab1}. }%
\label{fig6}%
\end{center}
\end{figure*}

As seen in Fig. \ref{fig6}, our evolutionary tracks describe well the advanced
evolutionary stage of these stars. Having obtained the mass of every star
using these tracks, it is possible to identify its evolutionary stage and
compare its CNO surface abundances with the observational data. This will be
described in Section 3.2.%

\begin{table}%
\caption{$a$: present work (M$_{RGB }$ is the mass on the RGB, including mass-loss), $b$: Tsuji08, $c$: Smith and Lambert (1985), $d$: Harris et al. (1988), $e$: Decin et al. (1997), $f$: El Eid (1994). The masses are given in solar units.  Note that the error on the mass by Tsuji08 increases for more massive stars, which may be related to the inaccuracy in the parallax measurements.}%
\begin{tabular}
[c]{|l|l|l|l|l|}\hline
Object(BS/HD) & M$_{initial}^{\textnormal{ }a}$ & M$_{RGB}^{\textnormal{ }a}$ & M $^{b}$ &
M(others)\\\hline\hline
$\delta$ Vir $(4910)$ & $1.2\pm0.2$ & $1.19\pm0.2$ & $1.4\pm0.3$ & $2.0$
$^{c}$\\\hline
$\alpha$ Tau $(1457)$ & $1.2\pm0.2$ & $1.20\pm0.2$ & $1.5\pm0.3$ &
$1.5^{\textnormal{ }c,d}$\\\hline
RRUMi $(5589)$ & $1.3\pm0.1$ & $1.15\pm0.1$ & $1.6\pm0.3$ & \\\hline
RZ Ari $(687)$ & $1.3\pm0.2$ & $1.14\pm0.2$ & $1.5\pm0.4$ & \\\hline
$\delta$ Oph $(6056)$ & $1.4\pm0.2$ & $1.39\pm0.2$ & $1.6\pm0.3$ & \\\hline
$\nu$ Vir $(4517)$ & $1.4\pm0.4$ & $1.39\pm0.4$ & $1.7\pm0.4$ & $2.0^{\textnormal{
}c}$\\\hline
$\tau^{4}$ Eri $(1003)$ & $1.8\pm0.3$ & $1.73\pm0.3$ & $2.0\pm0.4$ & \\\hline
$10$ Dra $(5226)$ & $1.8\pm0.3$ & $1.74\pm0.3$ & $2.1\pm1.8$ & \\\hline
$\beta$ Peg $(8775)$ & $1.8\pm0.3$ & $1.70\pm0.3$ & $2.2\pm0.3$ & $1.7$
$^{c,d,e,f}$\\\hline
$30g$ Her $(6146)$ & $1.8\pm0.3$ & $1.65\pm0.3$ & $2.0\pm0.6$ & $4.0^{c}%
$\\\hline
$\sigma$ Lib $(5603)$ & $2.0\pm0.2$ & $1.90\pm0.2$ & $2.2\pm0.5$ & \\\hline
R Lyr $(7157)$ & $2.0\pm0.2$ & $1.80\pm0.2$ & $2.1\pm0.5$ & \\\hline
$\mu$ Gem $(2286)$ & $2.0\pm0.5$ & $1.90\pm0.5$ & $2.3\pm0.5$ & $2.0$ $^{d.f}%
$\\\hline
OP Her $(6702)$ & $2.1\pm0.4$ & $1.90\pm0.4$ & $2.3\pm1.0$ & \\\hline
$\rho$ Per $(921)$ & $2.5\pm0.4$ & $2.40\pm0.5$ & $3.2\pm0.5$ & \\\hline
$\alpha$ Cet $(911)$ & $3.0\pm0.5$ & $2.96\pm0.5$ & $3.6\pm0.4$ & \\\hline
$\lambda$ Aqr $(8698)$ & $3.0\pm0.5$ & $2.96\pm0.5$ & $3.7\pm1.2$ & \\\hline
XY Lyr $(7009)$ & $4.0\pm1.0$ & $3.0\pm1.0$ & $3.7\pm1.5$ & \\\hline
$\delta^{2}$ Lyr $(7139)$ & $5.0\pm1.0$ & $4.5\pm1.0$ & $5.5\pm2.0$ & \\\hline
$\alpha$ Her $(6406)$ & $5.5\pm1.5$ & $4.0\pm1.0$ & $5.0\pm2.0$ & $7.0$ $^{f}%
$\\\hline
BS $6861(6861)$ & $5.5\pm1.5$ & $5.35\pm1.5$ & $6.3\pm4.0$ & \\\hline
\end{tabular}
\label{tab2}%
\end{table}%

The stellar masses are given in Table \ref{tab2}. For completeness, masses
obtained by other works for some stars are included
\citep{mai74,smi85,har88,dec97}. The error on the mass is determined from the
error bars on the observational M$_{\textnormal{bol}}$ and T$_{eff}$. Table
\ref{tab2} shows that our values are systematically lower than those by
Tsuji08. We attribute this mainly to two reasons:

(a) The stars are evolved to the stage where they are observed, that is, we do
not use any extrapolated tracks as done in Tsuji08.

(b) Mass-loss is taken into consideration, which becomes significant for the
more massive stars.

Moreover, it is clear from Table \ref{tab2} that our errors on the masses are
also lower. Calculating the tracks up to advanced stages helps to get better
evaluation of the masses of red giants.

\subsection{CNO Surface Abundances in Red Giants}

\subsubsection{Predictions of surface abundances with standard mixing}

This section presents the results for the surface CNO\ abundances of the
studied sample of stars. These are then compared to the CNO\ isotopic ratios
inferred from observations. The abundance profiles of the CNO isotopes and
those of H and He prior to the FDUP are shown in Fig. \ref{fig9}. In order to
study the variation of these abundance profiles as a function of the initial
stellar mass, two sample stellar masses are shown: $1.4$M$_{\odot}$ (Fig.
\ref{fig9}a) and $5$M$_{\odot}$ (Fig. \ref{fig9}b). Several features can be
identified in these figures:

\begin{figure*} 
\centering
\includegraphics[scale=0.418]{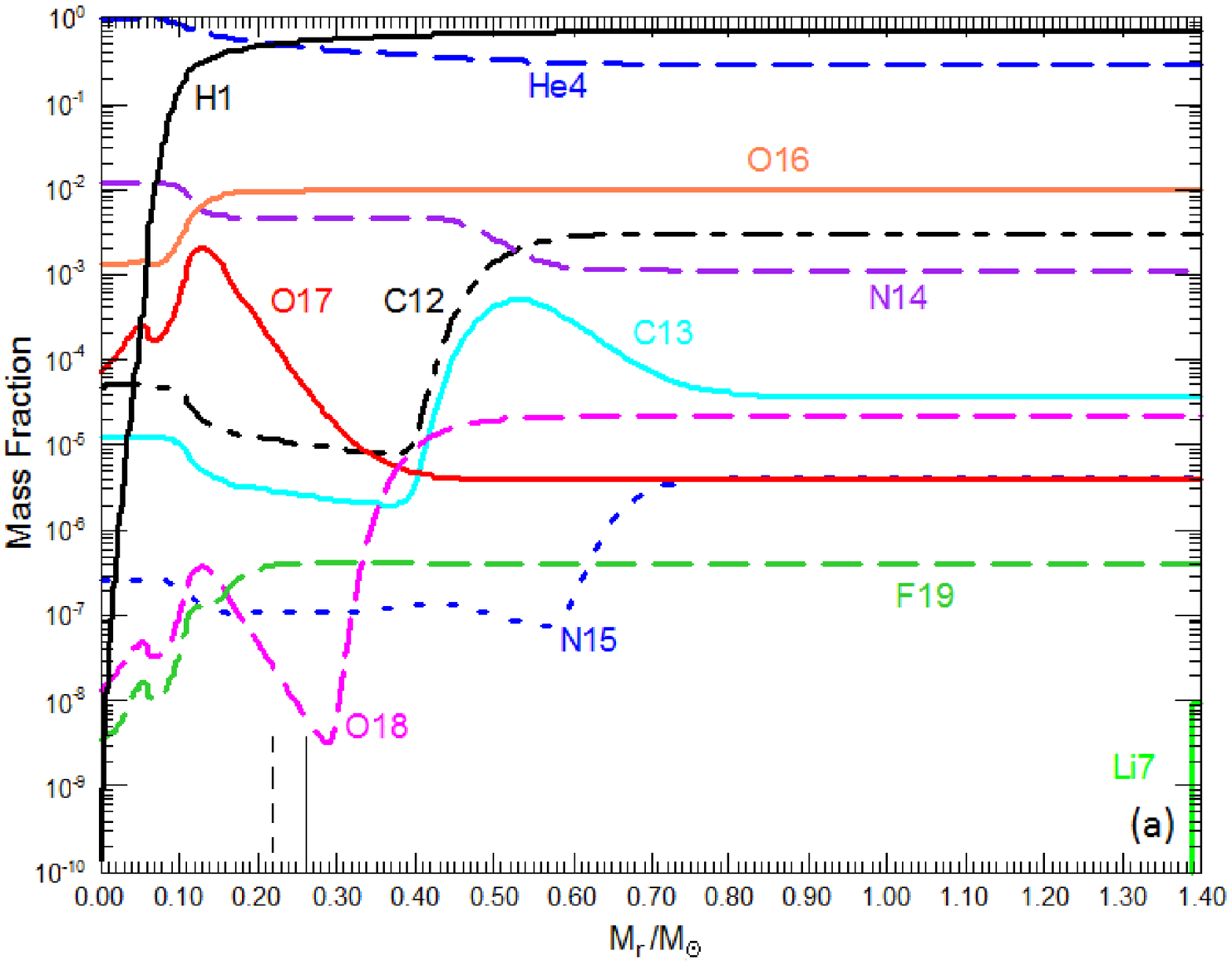}
\includegraphics[scale=0.418]{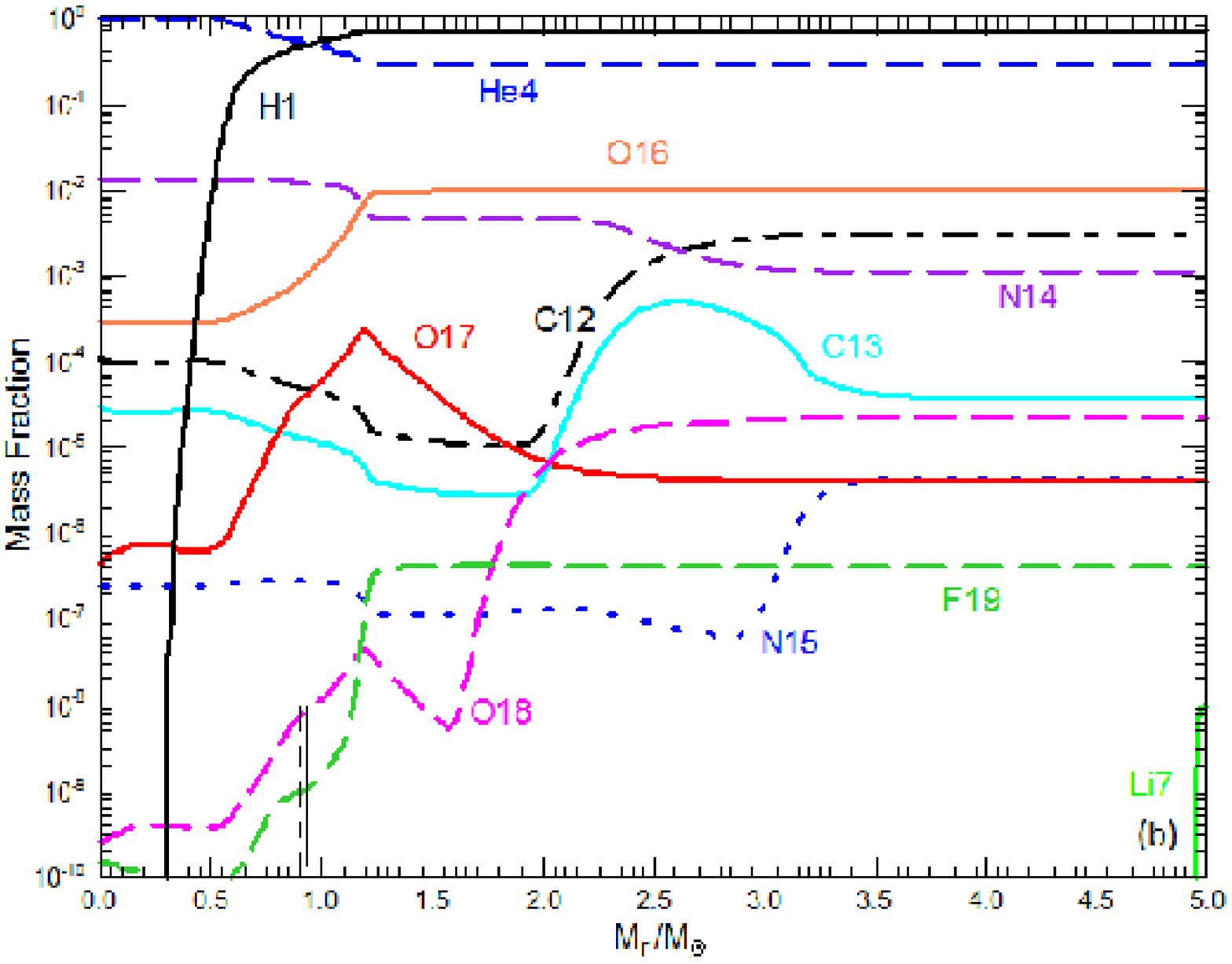}
\caption{Abundance profiles of H, He and several CNO elements after core H-burning in the stellar masses (a) 1.4M$_{\odot}$ and (b) 5M$_{\odot}$. The solid lines mark the deepest penetration in mass of the convective envelope, while the dashed lines mark that when envelope overshooting is invoked. See text for details.}
\label{fig9}
\end{figure*}

(a) The isotope $^{13}$C $\ $is produced near the middle of the star by the
CN\ cycle in all cases. This reflects the relatively low temperatures required
for the production of this isotope by the CN-cycle.

(b) The $^{17}$O profile is remarkable, showing a steep gradient in the
central region. This is because $^{17}$O is produced by the ON cycle which
requires higher temperatures to become effective.

(c) The isotope $^{18}$O is fragile and effectively destroyed by the $^{18}%
$O$($p$,\alpha)^{15}$N reaction$.$

These results are well known in the literature, but it is important to
understand the surface abundances resulting after FDUP\ and SDUP for different
stellar masses. The efficiency of FDUP in modifying the surface abundances is
related to the maximum penetration of the convective envelope on the RGB. In
Fig. \ref{fig9}, the solid vertical line marks the convective boundary as
determined by the Schwarzschild criterion and the dashed vertical line marks
the position of the boundary when extra mixing beyond the formal convective
boundary is considered (see Section 3.2.2 for details).

After FDUP, the change in the surface composition of a certain isotope depends
on the shape of its profile inside the star. For example, the peak of the
$^{13}$C profile is located near the middle part of the star, so that envelope
convection is able to smear out the profile, causing an increase in the
$^{13}$C surface abundance, or a decrease in the $^{12}$C$/^{13}$C ratio. For
the isotope $^{17}$O$,$ the situation is highly dependent on the stellar mass
because the main production of $^{17}$O is concentrated in the inner part.
Fig. \ref{fig9}a which shows the case of a $1.4$M$_{\odot}$ as an example,
illustrates that in low mass stars (M$\leqslant2$M$_{\odot}$) envelope
convection does not completely smear out the $^{17}$O peak. This makes the
$^{17}O$ profile sensitive to mixing in stars of M$\leqslant2$M$_{\odot}$, so
that any additional mixing below the envelope will increase its surface
abundance. This effect is less pronounced in stars with M $\geqslant
3$M$_{\odot}$ (Fig. \ref{fig9}b shows the case of a $5$M$_{\odot}$ as an
example in this mass range), where the $^{17}$O bump is fully engulfed by the
formal convective envelope, so extra mixing will not significantly alter its
surface abundance in this mass range.%

\begin{table*}%
\caption{M$_{RGB }$ (in solar units) and the surface number abundances after FDUP and SDUP (if any, listed in parenthesis): $a$: surface abundances with standard mixing (present work), $b$: Tsuji08, $g$: Harris and Lambert (1984b), $h$: H85: Harris et al. (1985), $i$: Smith and Lambert (1990), $j$: Decin et al. (2003), $k$: Maillard (1974), $l$: Hinkle et al. (1976).}%
\begin{tabular}
[c]{|l|l|l|l|l|l|l|l|l|}\hline
Object & M$_{RGB}$ & $^{16}$O$/^{17}$O$^{a}$ & $^{16}$O$/^{17}$O$^{b}$ &
$^{16}$O$/^{17}$O & $^{12}$C$/^{13}$C$^{a}$ & $^{12}$C$/^{13}$C$^{b}$ &
$^{12}$C$/^{13}$C & $^{14}$N$/^{15}$N$^{a}$\\\hline\hline
$\delta$ Vir & $1.19$ & $2370$ & $>2500$ &  & $32$ & $12.3\pm1.2$ &
$16\pm4^{i}$ & 551\\\hline
$\alpha$ Tau & $1.2$ & $2370$ & $>1000$ & $600_{+130}^{-150g}$ & $30$ &
$10.6\pm1.0$ & $10\pm2^{i,j}$ & 551\\\hline
&  &  &  & $525_{+250}^{-125g}$ &  &  & $9\pm\ 1^{g}$ & \\\hline
RRUMi & $1.15$ & $2006$ & $>2000$ &  & $30$ & $10.0\pm0.8$ &  & 598\\\hline
RZ Ari & $1.14$ & $2006$ & $607\pm48$ &  & $30$ & $7.9\pm0.8$ &  & 598\\\hline
$\delta$ Oph & $1.39$ & $1500$ & $387\pm68$ &  & $29$ & $11.1\pm0.9$ &  &
690\\\hline
$\nu$ Vir & $1.39$ & $1500$ & $>2000$ &  & $29$ & $8.7\pm1.3$ & $12\pm2^{i}$ &
690\\\hline
$\tau^{4}$ Eri & $1.73$ & $461$ & $687\pm14$ &  & $24$ & $12.4\pm0.3$ &  &
963\\\hline
$10$ Dra & $1.74$ & $461$ & $151\pm11$ &  & $24$ & $14.8\pm1.6$ & $12\pm3^{i}$
& 963\\\hline
$\beta$ Peg & $1.7$ & $462$ & $>2500$ & $1050_{+500}^{-250g}$ & $24$ &
$7.7\pm0.5$ & $8\pm2^{i}$ & 963\\\hline
&  &  &  & $\geq100^{g}$ &  &  & $5\pm3^{j}$ & \\\hline
$30g$ Her & $1.65$ & $462$ & $211\pm42$ & $675_{-175}^{+175}$ $^{h}$ & $24$ &
$12.5\pm1.1$ & $10\pm2$ $^{i}$ & 963\\\hline
$\sigma$ Lib & $1.9$ & $301$ & $>1500$ &  & $22$ & $7.5\pm0.3$ &  &
1126\\\hline
R Lyr & $1.8$ & $301$ & $368\pm44$ &  & $22$ & $6.4\pm0.3$ &  & 1126\\\hline
$\mu$ Gem & $1.9$ & $301$ & $798\pm73$ & $325_{+150}^{-75\textnormal{ }g}$ & $22$ &
$10.5\pm1.2$ & $13\pm2^{i}$ & 1126\\\hline
&  &  &  & $\geq100^{g}$ &  &  &  & \\\hline
OP Her & $1.9$ & $246$ & $329\pm31$ &  & $22$ & $11.3\pm1.2$ &  & 1103\\\hline
$\rho$ Per & $2.4$ & $234$ & $>1000$ &  & $22$ & $9.7\pm1.0$ & $15\pm2^{i}$ &
1307\\\hline
$\alpha$ Cet & $2.96$ & $318(317)$ & $586\pm47$ &  & 21($21)$ & $11.1\pm0.8$ &
$10\pm2^{j}$ & 1460(1501)\\\hline
$\lambda$ Aqr & $2.96$ & $318(317)$ & $>1000$ &  & 21($21)$ & $7.9\pm1.4$ &  &
1460(1501)\\\hline
XY Lyr & $3.0$ & 424($395)$ & $223\pm16$ &  & 21($20)$ & $15\pm0.4$ &  &
1528(1610)\\\hline
$\delta^{2}$ Lyr & $4.5$ & 400($378)$ & $465\pm41$ &  & 21($20)$ &
$16.2\pm1.5$ &  & 1503(1629)\\\hline
$\alpha$ Her & $4.0$ & 424($402)$ & $102\pm8$ & $180_{+70}^{-50g}$ & 21($20)$
& $11.1\pm0.7$ & $17\pm4^{g,l}$ & 1538/1642\\\hline
&  &  &  & $200_{+25}^{-25g}$ &  &  &  & \\\hline
&  &  &  & $\approx$ $450^{g}$ &  &  &  & \\\hline
&  &  &  & $450_{+50}^{-50k}$ &  &  &  & \\\hline
BS $6861$ & $5.35$ & 424($402)$ & $>1000$ &  & 21($20)$ & $48.5\pm2.9$ &  &
1538(1642)\\\hline
Initial &  & 2620 &  &  & 90 &  &  & 270\\\hline
\end{tabular}
\label{tab4}%
\end{table*}%

Table \ref{tab4} summarizes the values of $^{16}$O$/^{17}$O, $^{12}$C$/^{13}$C
and $^{14}$N$/^{15}$N after the FDUP and SDUP (if any, as we show later) in
the case of standard convective mixing, together with those inferred from
observations by Tsuji08 and other independent field star observations. The
stars in the sample of Tsuji08 are advanced in evolution, but did not
experience the third dredge up during the AGB\ phase. This is evident from the
carbon surface abundances (Tsuji 2014, private communication). Thus, the
comparison can be restricted to the effects of FDUP\ and SDUP only.

It is not easy to make a\ direct comparison between predicted surface
abundances and observations, since this comparison is model-dependent on
theoretical and observational grounds. Deriving the isotopic ratios
observationally involves several sources of uncertainty like the dispersion in
the ratios obtained from different lines, and inaccuracy in the atmospheric
model parameters. Systematic errors may also be present, such as the
uncertainty in the continuum position and departures from local thermodynamic
equilibrium (LTE) \citep{abi12}, in addition to the difficulties that are
inherent in the spectral analysis of very cool stars. The fact that
atmospheric values are model-dependent causes discrepancies between
observational results among different groups, and consequently, affect the
subsequent discussion \citep{ram14}. On the other hand, theoretical models are
also challenged by uncertainties in convective mixing, mass loss and nuclear
reaction rates, where standard FDUP models often face difficulties in
explaining carbon and oxygen surface abundances, particularly in low-mass stars. While
being vigilant to these limitations, a careful comparison is useful for the
sake of a better understanding.

\begin{figure}
\begin{center}
\includegraphics[scale=0.6]
{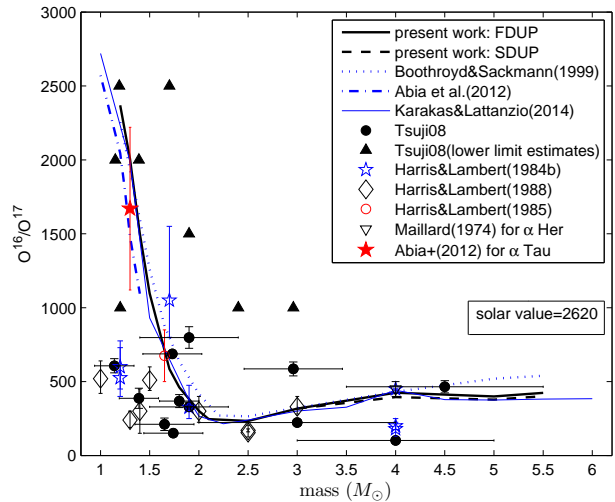}
\caption{Calculated $^{16}$O$/^{17}$O ratios for the 21 giants after FDUP
(solid line) and SDUP (dashed line) versus stellar mass. Calculation by
\citet{boo99}, \citet{abi12} and \citet{kar14} are also shown for comparison.
Observations are indicated as data points. The errors on the masses are as
listed in Table \ref{tab2}.}%
\label{fig8}%
\end{center}
\end{figure}

Fig. \ref{fig8} shows the $^{16}$O$/^{17}$O ratios as a function of stellar
mass, along with the theoretical predictions by \citet{boo99}, \citet{abi12},
\citet{kar14} and the available observations. The predicted $^{16}$O$/^{17}$O
shows a distinct behavior between high and low mass stars, which will be
discussed separately. For the higher masses (M$>$ $2$M$_{\odot})$, the
standard calculation predicts $^{16}$O$/^{17}$O ratios that fit the observed
range within the error bars on the stellar masses. The relative spread in the
observational values in this mass range cannot be explained by overshooting
since in these stars the bump in the $^{17}$O profile is fully mixed by the
formal convective envelope as shown in Fig. \ref{fig9}b, so overshooting will
not introduce a significant change to the standard predictions. Therefore, we
attribute this spread to the uncertainties in describing the properties of the
relatively cool red giants. This is a general feature of observationally
inferred data due to the above mentioned uncertainties in measuring faint
lines in their spectra which may result in considerably variant ratios for the
same object among different observations. We expect that the error bars of
these ratios are underestimated, particularly in stars like $\alpha$ Cet and
$\alpha$ Her whose ratio varies considerably among different works as shown in
Table \ref{tab4}.

In the case of the lower mass stars, the theoretical $^{16}$O$/^{17}$O is
higher than most data points obtained from observations. The results of
standard calculation by other groups shown in the figure indicate a similar
behavior. It is noted that the high observational $^{16}$O$/^{17}$O values in
the low mass stars like $\alpha$ Tau, $\nu$ Vir, $\sigma$ Lib, $\beta$ Peg,
$\delta$ Vir, and RR UMi, which are denoted by up-pointing triangles in Fig.
\ref{fig8}, are estimations due to the uncertainties in measuring faint lines
like $^{12}$C$^{17}$O in relatively cool atmospheres, where at such low
temperatures the absorption lines are strong causing severe blending by
several weak lines and introducing uncertainties (Tsuji08). Due to such
complications, weak lines couldn't be measured at all for $^{12}$C$^{17}$O in
these stars, and $^{16}$O$/^{17}$O is not well determined. Thus, these
observational data points are not reliable to compare to our calculation. Our
discussion is based on the low observational ratios in low mass stars which
are more definitive, with well-determined error bars.

The overproduction of $^{16}$O$/^{17}$O in low mass stars in our model shows
that more $^{17}$O needs to be mixed to the surface in order to lower the
$^{16}$O$/^{17}$O ratio. This may be achieved by extra mixing or overshooting
at the bottom of the convective envelope. In order to constrain this
overshooting, a better understanding is required for the role of first and
second dredge up for the whole range of masses under consideration.%

It is known that after the star leaves the main sequence, FDUP alters
significantly the surface abundances as it enriches the envelope with $^{4}%
$He, $^{13}$C, $^{17}$O and $^{14}$N and reduces its $^{12}$C, $^{15}$N and
$^{18}$O abundance. Every star that evolves to the AGB experiences the FDUP
episode, and starts its early AGB phase with a sharp composition discontinuity
at the point of maximum penetration of the FDUP. The H-burning shell in
low-mass stars represents an entropy barrier that prevents any deeper
penetration of the convective envelope, and thus no further change in the
surface abundances takes place. However, for solar metallicity stars of masses
above $3$-$4$M$_{\odot}$, the gravitational energy release due to the
contracting core and the increased energy flux from the burning He-shell lead
to an expansion so that the H-shell is pushed outwards in mass to low
temperatures causing a temporary extinction in the H-burning shell. This
situation, coupled to the drop in the temperature of the expanding layers and
the increase in the opacity \citep{ib83}, causes convection to deepen in a
second dredge up (SDUP) event and mixes out the composition discontinuity left
over by the FDUP. Therefore, SDUP introduces further changes in the surface
abundances of stars of masses $\gtrsim4$M$_{\odot}$.%

\begin{figure}
\begin{center}
\includegraphics[scale=0.6]
{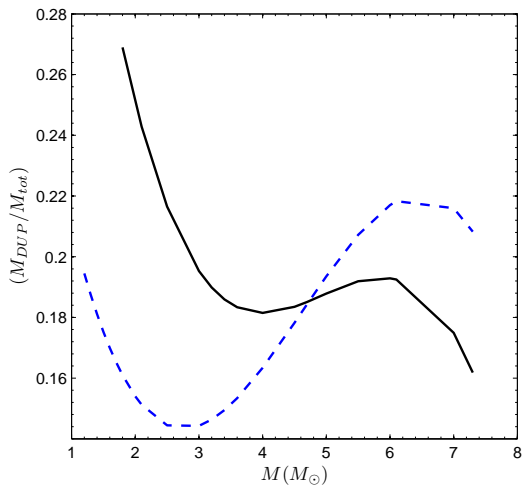}
\caption{Fractional mass reached by the convective envelope at FDUP (dashed
line) and SDUP (solid line). SDUP is deeper that FDUP only in models of
stellar masses $\geqslant5$M$_{\odot}$. The mass on the y-axis is given as the
mass reached by the dredge up over the total initial stellar mass, in solar
units.}%
\label{fig10}%
\end{center}
\end{figure}

Fig. \ref{fig10} shows the maximum penetration of the convective envelope at
FDUP and SDUP as a function of the initial stellar mass. It is clear from the
figure that for M$<$ (4-$5)$M$_{\odot}$, SDUP doesn't penetrate deeper than
the FDUP, and thus, it does not introduce significant change to the surface
composition. This is in agreement with \citet{kar14}, for their solar
metallicity case. Their Fig. 7 exhibits similar general features and also
indicates a deepest penetration of FDUP in $\approx2.5$M$_{\odot}$ stars. For
stars of M%
$>$%
$6$M$_{\odot},$ the expansion is strong and convection extends deeper inwards.
Since low mass stars do not experience SDUP, this implies that their observed
surface abundances on the AGB phase are actually \textquotedblleft
inherited\textquotedblright\ from the RGB phase.\ Therefore, one way to
account for the discrepancy between calculated and observationally inferred
values of $^{16}$O$/^{17}$O in these stars is to invoke extra mixing below the
convective envelope on the RGB phase.
\subsubsection{Predictions of surface abundances with extra mixing}

\begin{figure}
\begin{center}
\includegraphics[scale=0.6]
{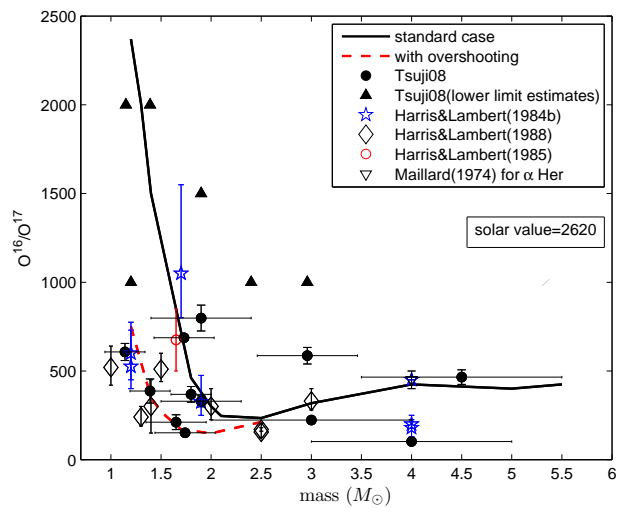}
\caption{The $^{16}$O$/^{17}$O ratios obtained in the standard case and in the
case of envelope overshooting with $f$=0.125. Several observations are shown
for comparison.}%
\label{fig12}%
\end{center}
\end{figure}

Overshooting is applied as outlined in Section 2.1 and we find that $f=0.125$
provides the best estimation for its efficiency in low mass stars. Fig.
\ref{fig12} shows the $^{16}$O$/^{17}$O ratios of our star sample in the
standard case and with overshooting. The low observational $^{16}$O$/^{17}$O
in low mass stars can now be fitted quite well within the error bars on the
stellar masses. It is noted how overshooting has a minor effect in stars of M
$\geqslant3$M$_{\odot}$. This is expected due to the shape of the $^{17}$O
abundance profile in these stars as discussed earlier in connection with the
shape of the $^{17}$O profile in these stars. The calculated and
observationally inferred values for these stars are generally in a good agreement.

\begin{figure*} \centering
\includegraphics[scale=0.59]
{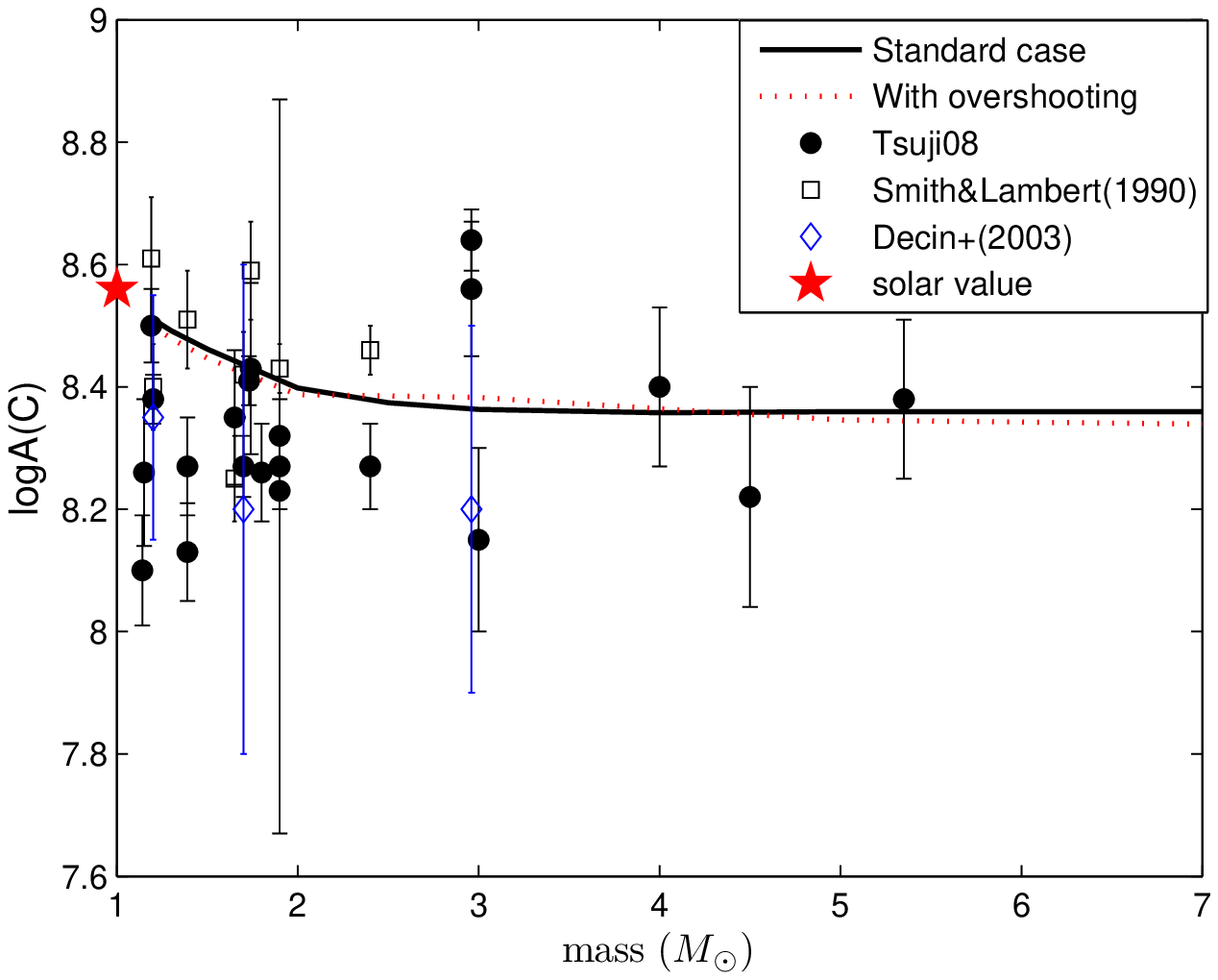}
\includegraphics[scale=0.59]
{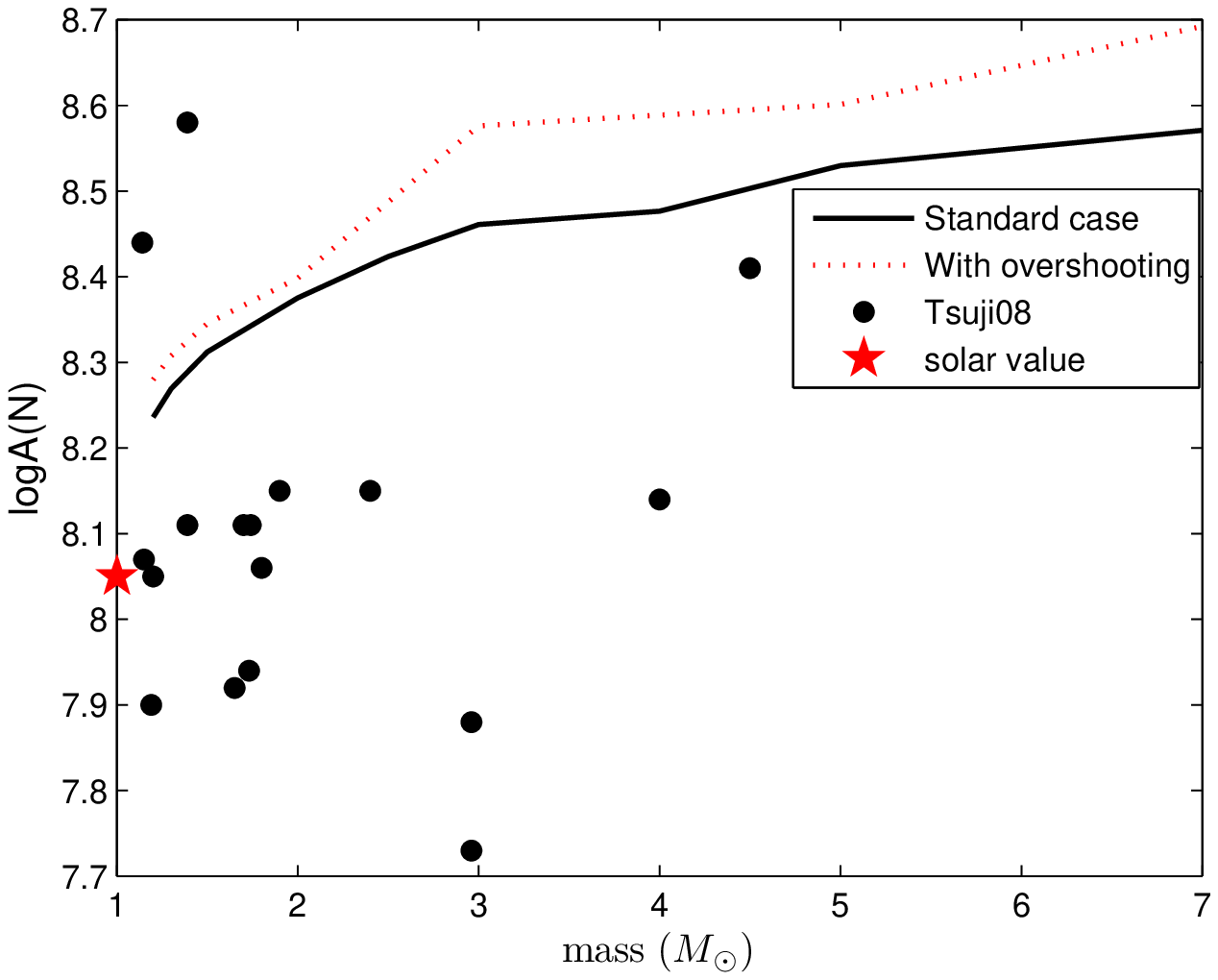}
\caption{The predicted surface abundances of (a) carbon and (b) nitrogen with observations as shown. The initial solar values of carbon and nitrogen are 8.5601 and 8.0499, respectively and are indicated by a star symbol.}\label{c&n}%
\end{figure*}

Fig. \ref{c&n} shows the predicted carbon and nitrogen abundances in the
standard case and with overshooting. Observations are also shown for
comparison. A well-known problem arises in explaining the surface carbon
abundances in stars of M$\leqslant2$M$_{\odot}$, where our extra mixing
treatment cannot explain the low carbon observed in these giants. Calculations
using standard mixing by \citet{boo99} and \citet{kar14} show a similar
problem. A different non-standard mixing process seems to be required in
low-mass stars ($<2$M$_{\odot})$ which can reduce the abundance of $^{12}$C by
mixing it to the hotter regions of the H-burning shell, allowing for some
nuclear processing. Such mixing process is linked to the long evolutionary
time of the low-mass stars while ascending the RGB. As discussed in some
details in \citet{kar14}, the so-called \textquotedblleft thermohaline
mixing\textquotedblright\ or \textquotedblleft salt-finger
instability\textquotedblright\ seems to be one possible description of this
extra mixing in low-mass red giant stars. This is still an unsettled problem
in stellar modeling. Nitrogen abundances, on the other hand, show a scatter in
the observed data. This will be discussed in connection with the $^{14}%
$N$/^{15}$N ratios in Section 3.3.2. In general, Fig. \ref{c&n}b shows that the
observations pose no serious contradiction with the values predicted by the
stellar models which, on average, reasonably reproduce the observationally
inferred data.%

%

\subsection{Effect of modified nuclear reaction rates}

\subsubsection{The $^{16}$O$/^{17}$O ratio}

\begin{figure*} \centering
\includegraphics[scale=0.58]
{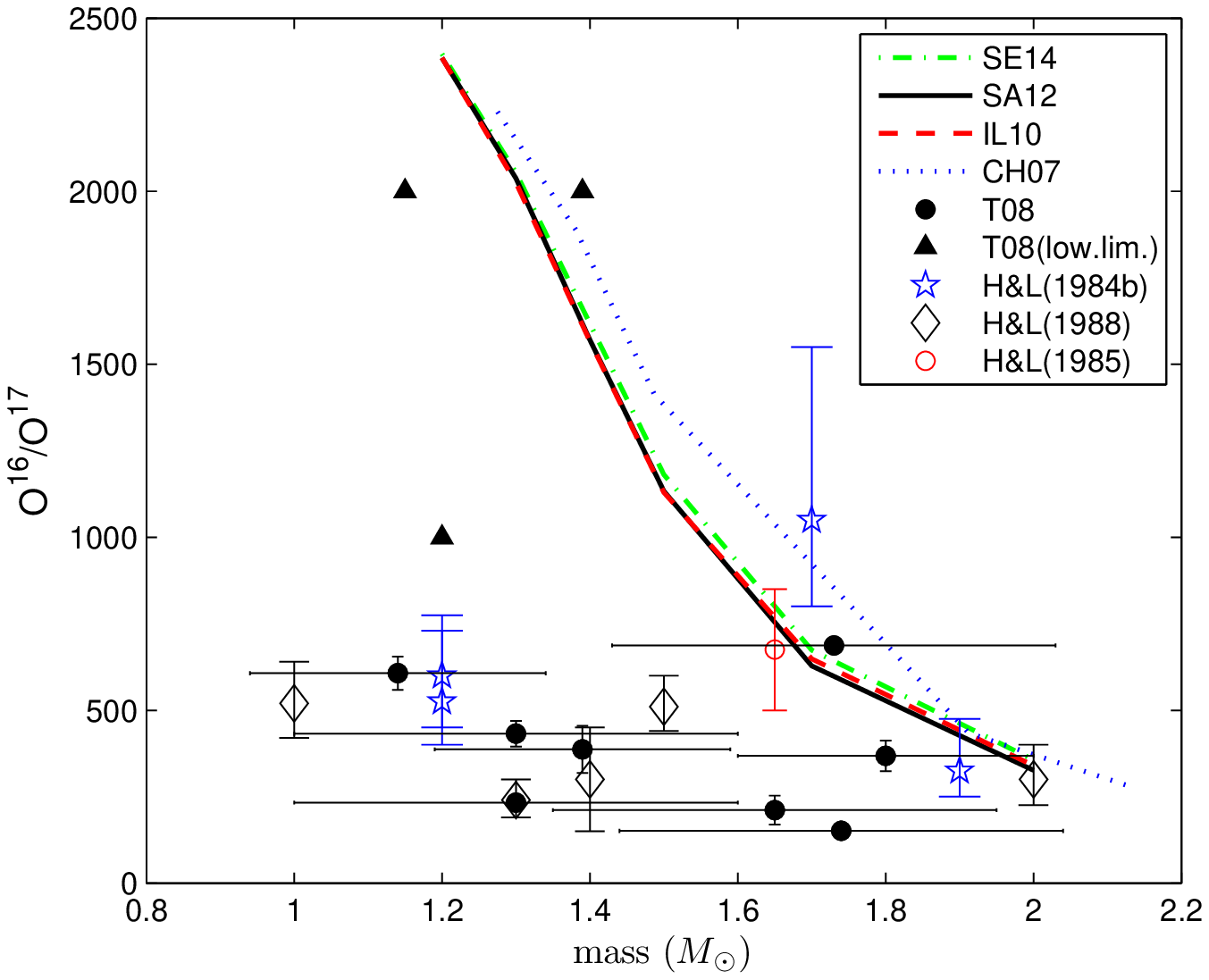}
\includegraphics[scale=0.58]
{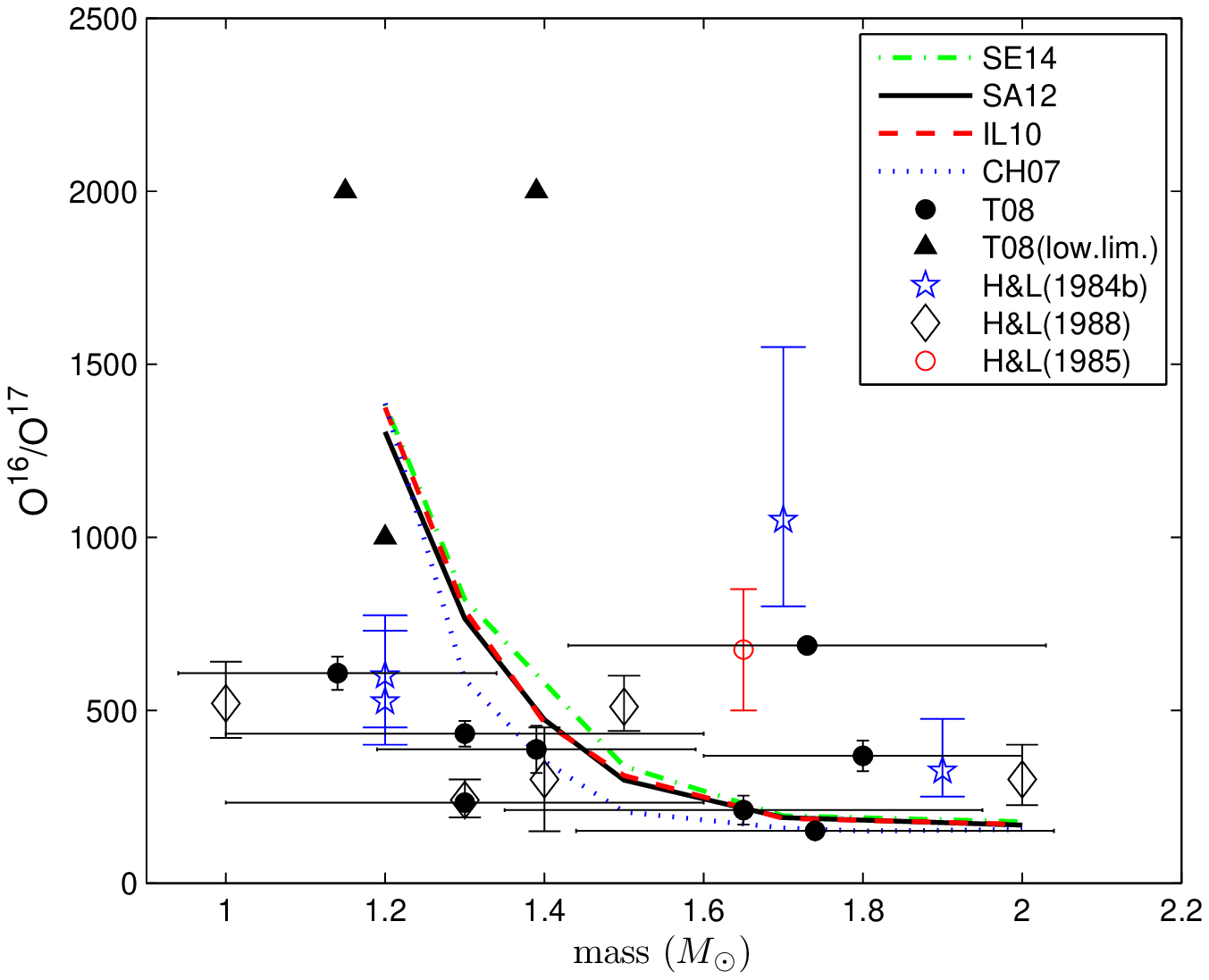}
\caption{The $^{16}$O/$^{17}O$ ratios calculated for the 21 giants, with four different sets of compilations of relevant reaction rates (see text for details). Left panel shows the case with standard mixing and the right panel is that with overshooting ($f=0.125$). Observational data are also included.}\label{fig14}
\end{figure*}

In order to investigate the effect of the $^{17}$O production and destruction
reaction rates on the $^{16}$O$/^{17}$O ratios, four different evaluations of
the proton-capture reactions $^{16}$O(p$,\gamma)^{17}$F$,$ $^{17}$%
O(p$,\gamma)^{18}$F and $^{17}$O(p$,\alpha)^{14}$N are used. In particular, we
use the compilations by \citet{ser14} (SE14), \citet{sa13} (SA13),
\citet{il10} (IL10) and \citet{ch07} (CH07). The $^{16}$O$/^{17}$O ratios
obtained are shown in Fig. \ref{fig14} for both cases: standard mixing (Fig.
\ref{fig14}a) and envelope overshooting with $f=0.125$ (Fig. \ref{fig14}b).
The four sets of rates give very similar $^{16}$O$/^{17}$O values in the
considered mass range. None provides a reasonable agreement between model
predictions and observations unless overshooting is included. This consistency
shows that the existing discrepancy cannot be removed without invoking deeper
mixing. Fig. \ref{fig14}b shows a better fit of the observed data within the
error bars in the low mass stars.

\begin{figure}
\begin{center}
\includegraphics[scale=0.61]
{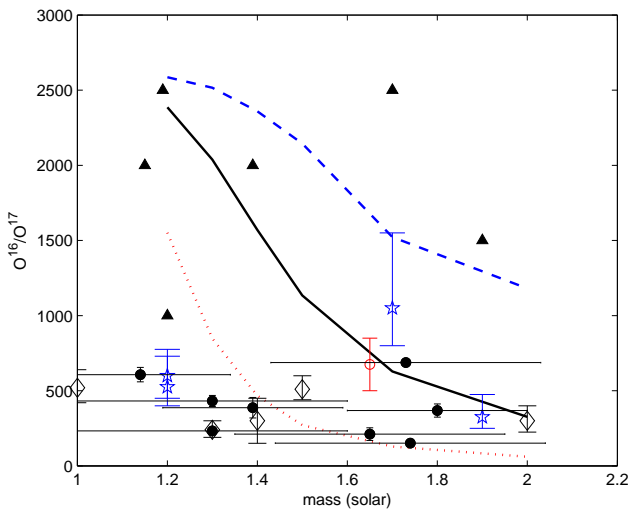}%
\caption{The SA13 set of rates in the standard case. Shown are the results
with the recommended set of rates (solid line), the high rates (dotted) and
low rates (dashed). Observational data are the same as in Fig. \ref{fig14}. }%
\label{fig16}%
\end{center}
\end{figure}

The effect of the reaction rates uncertainties is also worth exploring. Since
the SA13 compilation is based on a Monte Carlo simulation and the rates have
statistically well-defined uncertainties \citep{lon10,lon12,ili14}, the
$^{16}$O$/^{17}$O ratios are calculated using the recommended, high and low
rates, where the rate boundaries correspond to a 95\% coverage probability.
The rates uncertainty has a very minor effect on\ the tracks during H-shell
burning but none along the RGB, and hence it does not affect our mass
determination. However, the $^{16}$O$/^{17}$O ratios show a larger sensitivity
to these uncertainties, as shown in Fig. \ref{fig16}. It is found that the
observational data can be better explained when the whole range of uncertainty
on the involved reaction rates is considered since the discrepancy between
predictions and observations becomes less pronounced. On the other hand, the
$^{12}$C$/^{13}$C ratios are found to be immune to the uncertainties on the
considered rates. The difference in $^{12}$C$/^{13}$C obtained with the high
rates and low rates does not exceed 3\%.

\subsubsection{The $^{14}$N$/^{15}$N ratio}

The surface $^{14}$N$/^{15}$N ratio is worth considering in connection with
the latest evaluation of the $^{14}$N(p$,\gamma)^{15}$O rate \citep{mar11}.
This ratio is calculated for the mass range under consideration after FDUP and
SDUP. It is found that our values are higher by $\sim$ 20\% relative to those
by \citet{eid94} in stars of masses $\geqslant3$M$_{\odot}$. This is expected
and due to the NACRE $^{14}$N(p$,\gamma)^{15}$O rate \citep{ang99}  used in that work, which
is almost a factor of 2 higher than the revised rate at stellar temperatures
(see \citet{hal12} for the expression of this rate and explicit discussion).
However, our $^{14}$N$/^{15}$N is about 12\% lower in the low-mass stars.

It is quite unfeasible to verify our findings since $^{14}$N$/^{15}$N ratios
are difficult to measure in RGB stars from CN lines because they are too weak
even in very high resolution spectra. In fact, the $^{14}$N$/^{15}$N ratio is
difficult to measure directly for molecules containing carbon \citep{rou15}.
On the other hand, stars in the early AGB phase are O-rich and then their
spectra would be dominated by oxide molecules and not by C-bearing molecules.
Oxide molecules like NO cannot be observed in the visual or near-IR region in
these stars (Carlos Abia, private communication). Observations in
radio-wavelengths can in principle detect some N-bearing molecules from which
$^{14}$N$/^{15}$N may be derived, but in this case one would probably be
looking at the circumstellar envelope of the star not the photosphere.
Circumstellar N ratios might be affected by the incoming UV radiation from the
ISM triggering non-kinetic equilibrium chemistry and thus, might not represent
the stellar photospheric ratios \citep{ped12}. Theoretical predictions of the
$^{14}$N$/^{15}$N ratios, particularly the changes induced by the revision of
the $^{14}$N(p$,\gamma)^{15}$O reaction rate may benefit from future
observations or one may resort to the chemical analysis of pre-solar grains
originating from the envelopes of late AGB\ stars.

\section{Conclusions}

A sample of observed RGB and early AGB stars was considered and their masses
were obtained using extended evolutionary tracks. This was done without the
need of extrapolating the evolutionary tracks at lower effective temperatures
as it has been done in the work by Tsuji08. The present evolutionary tracks
include the effect of mass-loss, which becomes important during the red giant
phases, especially for the higher stellar masses under consideration. This
investigation includes an analysis of the physical conditions in the
overshooting region and the effect of this overshooting on the abundance
yields. We find that overshooting is needed to reconcile observational oxygen
abundances with model predictions particularly in low mass red giants.
Although Tsuji08 has interpreted the discrepancies between the predicated and
observational abundances as an effect of extra mixing, he did not provide
models including this extra mixing to see how the discrepancies may be
understood. We showed that extra mixing based on the mixing of chemical
elements by diffusion explains reasonably the observationally inferred oxygen
isotopic ratios. It is important to realize the challenges facing such
observations and the uncertainties involved in the available data. In this
regard, the spread in the observational data in the low and high mass ranges
was discussed in connection with the inherent difficulties in analyzing the
spectra of these relatively cool stars and the uncertainties involved in
measuring faint lines like $^{12}$C$^{16}$O. Our overshooting treatment
cannot, however, explain the low surface carbon abundances in low mass stars.
Another mixing mechanism seems to be required during the long evolutionary
time needed for the low-mass stars to ascend the red giant branch.

Furthermore, the present calculations were carried out using recent
determinations of proton-capture rates which have reliable statistical error
bars. This allows us to draw conclusions on the uncertainties involving CNO
surface abundances. In particular, the effect of recent evaluations of the
reaction rates on the production and destruction of $^{17}$O was explored$.$
The experimentally suggested uncertainty of these rates provides a better fit
of the $^{16}$O$/^{17}$O observed in low mass stars yet does not exclude the
need to invoke overshooting. Additional mixing beyond the convective envelope
as determined by the Schwarzschild criterion is found to be necessary to
better explain the observational $^{16}$O$/^{17}$O\ surface abundances,
especially in low mass stars. Moreover, the effect of $^{14}$N(p$,\gamma
)^{15}$O rate on the $^{14}$N$/^{15}$N ratios was studied in the considered
mass range.

As a final remark, our approach in the present study was to consider a sample
of red giant stars to see how far standard calculation predictions agree with
observations. A comprehensive comparison between stellar models and
observations based on the analysis of the effect of extra mixing was presented
and linked to the uncertainties in key nuclear reaction rates affecting the
CNO abundances in red giants. Comparing theoretical predictions of stellar
models to available observations is required in order to constrain
parametrized approaches in determining the efficiency and extension of mixing
at convective boundaries. Future multi-dimensional simulations of convection
may introduce an improved local description of this mixing and provide insight
towards a better understanding of the physical processes involved.

This work was supported by the Lebanese National Council for Scientific
Research (CNRS) under grant award number 11179-102899. The authors wish to
thank Christian Illiadis and Richard Longman for providing the nuclear
reaction rates and for their valuable input and discussions on the manuscript.
The authors also acknowledge the anonymous referee for the constructive
comments and suggestions that helped improve the quality of this work.

\end{document}